\documentclass[twocolumn,aip,reprint,superscriptaddress]{revtex4-2}

\usepackage{amsmath,amsfonts}
\usepackage{graphicx}

\usepackage{booktabs}
\usepackage{dcolumn}

\usepackage{bm}
\usepackage{physics}
\usepackage{xcolor}

\newcommand{\im}{\textup{i}}
\newcommand{\calbm}[1]{\bm{\mathcal{#1}}}
\newcommand{\one}{\mathrm{1}}
\newcommand{\unitval}[2]{#1\,#2}
\newcommand{\wavenumber}{\textup{cm}^{-1}}
\newcommand{\kelvin}{\mathrm{K}}
\newcommand{\nl}{\nonumber \\}

\begin{document}


\title{Spin-lattice relaxation with non-linear couplings: Comparison between Fermi's golden rule and extended dissipaton equation of motion}

\author{Rui-Hao Bi}
\affiliation{Department of Chemistry, School of Science and Research Center for Industries of the Future, Westlake University, Hangzhou, Zhejiang 310024, China}

\author{Yu Su}%
\affiliation{Hefei National Research Center for Physical Sciences at the Microscale, University of Science and Technology of China, Hefei, Anhui 230026, China}
\author{Yao Wang}%
\affiliation{Hefei National Research Center for Physical Sciences at the Microscale, University of Science and Technology of China, Hefei, Anhui 230026, China}

\author{Lei Sun}
\affiliation{Department of Chemistry, School of Science and Research Center for Industries of the Future, Westlake University, Hangzhou, Zhejiang 310024, China}
\affiliation{Institute of Natural Sciences, Westlake Institute for Advanced Study, Hangzhou, Zhejiang 310024, China}
\affiliation{Key Laboratory for Quantum Materials of Zhejiang Province, Department of Physics, School of Science and Research Center for Industries of the Future, Westlake University, Hangzhou, Zhejiang 310024, China}

\author{Wenjie Dou}%
\email{douwenjie@westlake.edu.cn} 
\affiliation{Department of Chemistry, School of Science and Research Center for Industries of the Future, Westlake University, Hangzhou, Zhejiang 310024, China}
\affiliation{Institute of Natural Sciences, Westlake Institute for Advanced Study, Hangzhou, Zhejiang 310024, China}
\affiliation{Key Laboratory for Quantum Materials of Zhejiang Province, Department of Physics, School of Science and Research Center for Industries of the Future, Westlake University, Hangzhou, Zhejiang 310024, China}

\date{\today}

\begin{abstract}
Fermi's golden rule (FGR) offers an empirical framework for understanding the dynamics of spin-lattice relaxation in magnetic molecules, encompassing mechanisms like direct (one-phonon) and Raman (two-phonon) processes. These principles effectively model experimental longitudinal relaxation rates, denoted as $T_1^{-1}$. However, under scenarios of increased coupling strength and nonlinear spin-lattice interactions, FGR's applicability may diminish. This paper numerically evaluates the exact spin-lattice relaxation rate kernels, employing the extended dissipaton equation of motion (DEOM) formalism. Our calculations reveal that when quadratic spin-lattice coupling is considered, the rate kernels exhibit a free induction decay-like feature, and the damping rates depend on the interaction strength. We observe that the temperature dependence predicted by FGR significantly deviates from the exact results since FGR ignores the {higher order effects and the} non-Markovian nature of spin-lattice relaxation. Our methods can be {easily extended to study} other systems with nonlinear spin-lattice interactions and provide valuable insights into the temperature dependence of $T_1$ in molecular qubits {when the coupling is strong}.
\end{abstract}

\maketitle

\section{\label{sec:intro} Introduction}
Quantum information technologies, such as quantum computation and quantum sensing, \cite{QuantumComputaiton-McArdle2020,QuantumSensingRev-Degen2017} rely on the efficient implementation of qubits. \cite{MOFMatRev-Yu2020,MOFasHost-Oanta2022} Among various implementations, including superconducting Josephson junctions, \cite{Josephson-Blais2021} photons,\cite{Photon-Pelucchi2021} and semiconducting quantum dots, \cite{SC_Qdots-Kloeffel2013} paramagnetic coordination complexes, i.e., the molecular qubits, are standing out due to their inherent synthetic versatility {and scalability}. \cite{Mol-Arom2012, Mol-Coronado2019, Atzori2019} {In particular, it is practical to carefully design these qubits, and reproduce them in large amount effortlessly.}

{The basic idea for the molecular spin qubits is based on the Zeeman effect, the} unpaired electrons {in molecules can yield} quantized spin states under an applied magnetic field. {Here, the spin Hamiltonian parameters can be optimized via synthetic chemistry techniques.} \cite{Atzori2019} Moreover, metal-organic frameworks (MOFs) and other materials have been utilized to assemble or polymerize these molecules into periodic arrays, enabling chemists to meticulously tune the topology and inter-qubit distance. \cite{ChemTune-Yu2019,molecular-design-Yu2020,molecular-design-Moisanu2023} {Overall, the molecular approach to  qubits promises a bottom-up chemical design, which}
provides a systematic {roadmap} {for the optimization of a single qubit and the scaling up of quantum applications.} \cite{MOFMatRev-Yu2020, MOFasHost-Oanta2022, Design-Graham2017, GeneralRev-Wasielewski2020}

{One of the current focuses in molecular spin qubits is to design robust qubits with a long coherence time, which is} essential for quantum error correction and fault-tolerant quantum computation. \cite{Design-Graham2017, GeneralRev-Wasielewski2020} {Through chemical designs, such as minimizing the nuclear magnetic moment, \cite{minimize-spin-Zadrozny2015} restricting certain vibrational or rotational modes, \cite{vib-kazmierczak_impact_2021,rot-qiu_enhancing_2023,vib-Kazmierczak2024} and mixing of the qubits spin state, \cite{clock-bayliss_enhancing_2022, clock-shiddiq_enhancing_2016}} {record} long coherence times {have been achieved.}  {Here, one of the goals is to design molecular qubits that have a long coherence time at higher temperatures, such as liquid nitrogen temperature or even room temperature, as it allows for broader applications for quantum technologies.}

{At high temperatures, the quantum coherence time is mainly controlled by the longitudinal relaxation. In particular}, the {decoherence} of a {molecular spin} qubit is constrained by both {the} spin phase coherence ($T_2$) and {the} longitudinal relaxation ($T_1$) times. \cite{T2T1-Jarmola2012, Design-Graham2017}  In applications at low temperatures, $T_1$ times notably surpass $T_2$, thus exerting minimal influence on qubit performance. \cite{T2T1-Jarmola2012} However, as temperature rises, $T_1$ becomes the primary factor limiting qubit performance. \cite{Design-Graham2017} Therefore, a comprehensive understanding of the temperature dependence of $T_1$ is crucial for designing qubits suitable for operation under high temperatures.

The key physics behind the $T_1$ relaxation time and its temperature dependency are the spin-lattice interactions, i.e., the coupling between the spin system with molecular phonon modes. To understand the relaxation dynamics, the simplest and most widely used models are based on Fermi's golden rule (FGR). {\cite{VanVleck1940,Orbach-1961,torchia_spin-lattice_1982,Egorov1995-multiphonon}} These {classic} models are often used to fit the temperature dependence of $T_1$ to empirical ``processes'', such as the direct (one-phonon) process and Raman (two-phonon) process. \cite{EPR-Goldfarb2018} {In addition to the relaxation rates, master equation theories were developed to study time evolution of the dissipative dynamics in spin-lattice relaxation. \cite{Redfield-1957,tjon_quantum_1964,emid_master_1973} Notably, the master equation approaches are being actively incorporated in \emph{ab initio} simulations and have seen great success when compared with experimental data.\cite{lunghi_role_2017, lunghi_limit_2020, lunghi_multiple_2020, lunghi_toward_2022, garlatti_critical_2023, nabi_accurate_2023}}
However, these empirical models {and master equation simulations} are based on the assumption of weak {system-bath couplings} and {a} Markovian bath, and thus are not always valid. 
{
In this work, we explore what happens when these approximations break down.} Although there have been exact simulations, such as the hierarchical equations of motion (HEOM) approach, \cite{HEOM-Tanimura1989,Perspective-Tanimura2020,takahashi_open_2020} and generalized master equations which take into account non-Markovian baths; \cite{chang1993NonMarkovianPopulationPhase, blanga1996MemoryEffectsSpin} these methods all focus on linear spin-lattice interactions, neglecting non-linear couplings. \cite{EDEOM2-Xu2018} 
As a result, there is a lack of studies to explore the effects of non-linear spin-lattice interactions and stronger coupling strengths on the temperature dependence of $T_1$.

In this study, we address this gap by comparing the FGR predictions with the numerically exact results from the extended dissipaton equation of motion (DEOM) formalism. \cite{EDEOM1-Xu2017,EDEOM2-Xu2018} We demonstrate that the temperature dependence of $T_1$ can significantly deviate from the FGR predictions when quadratic spin-lattice interactions are considered. Our results suggest that the temperature dependence of $T_1$ is sensitive to the interaction strength and the non-Markovian nature of the spin-lattice relaxation. Our findings provide valuable insights into the temperature dependence of $T_1$ in molecular qubits, and can be readily applied to other systems with non-linear spin-lattice interactions.

The rest of the paper is organized as follows. In Sec.~\ref{sec:theory}, we introduce the spin-boson model for spin-lattice relaxation, Fermi's golden rule, and the extended DEOM formalism. In Sec.~\ref{sec:results}, we present the results of our simulations and compare the FGR predictions with the numerically exact results. Finally, in Sec.~\ref{sec:conclusion}, we conclude our study and discuss the implications of our findings.

\section{\label{sec:theory} Theory}
\subsection{\label{subsec:model} Spin-boson model for spin-lattice relaxation}
{First, we consider a simplified model describing a spin-$\frac{1}{2}$ system interacting with a vibrational environment.} The total Hamiltonian reads \cite{chang1993NonMarkovianPopulationPhase, Egorov1995-multiphonon, Lunghi2023, TwoLevelDissipative-Leggett1987} (We set $\hbar$ as unit.)
\begin{align}\label{eqn:H}
    H_\text{T} &= \frac{\Delta}{2} \hat\sigma_z + \frac{1}{2}\sum_k \omega_k (\hat{p}_k^2 + \hat{q}_k^2) + H_1\nl 
    &= H_\text{S} + H_\text{B} + H_1
\end{align}
with $\Delta$ being the Zeeman energy induced by the static external field and $\{\sigma_i\}_{i=x,y,z}$ the Pauli matrices. {Here, the second term $H_\text{B}$ represents the phonon environment of the spin, where $\omega_k$ is the vibrational frequency of the $k$-th bath oscillator. The spin relaxation dynamics will be determined by how
these vibration modes couple to the spin system}. {Here, we consider} the spin-lattice {interaction} Hamiltonian {that can be represented in this specific form as}
\begin{gather}
H_1 = \hat{Q}_\text{S} (\alpha_0 + \alpha_1 \hat{x}_\text{B} + \alpha_2 \hat{x}_\text{B}^2 + \cdots), \label{eqn:H1}
\end{gather}
where $\hat{Q}_\text{S}$ is some spin--subspace operator and {$\hat{x}_\text{B}$ denotes a collective lattice mode consisted of linear combination of the bath oscillators, given by $\sum_k c_k \hat{q}_k$. Here, it is assumed the interaction between the spin and the collective lattice mode can be written as a Taylor expansion, and $\{\alpha_0,\alpha_1,\alpha_2, \dots\}$ denote the polynomial coefficients. In a concrete simulation, the collective mode coefficients and the polynomial coefficients need to be determined. Ref.~\onlinecite{Lunghi2023} demonstrated that these parameters could be obtained}  from \emph{ab initio} calculations.
%


{To proceed, we assume the collective mode consists of a continuum of oscillators, as in the cases of condensed phase, which can be characterized by a continuous spectral density function},
\begin{equation}\label{eqn:spectralfunction}
    J(\omega) = \frac{\pi}{2} \sum_k c_k^2 \delta(\omega - \omega_k), \quad \text{for $\omega>0$,}
\end{equation}
and $J(-\omega) = -J(\omega)$. The spectral density relates to the bare--bath response function $\chi_{\rm B}(t) \equiv i\langle[\hat x_{\rm B}(t),\hat x_{\rm B}(0)]\rangle_{\rm B}$ via 
\begin{align}
    J(\omega) = \frac{1}{2{\rm i}}\int_{-\infty}^{\infty}\!\!{\rm d}t\,e^{\im\omega t}\chi_{\rm B}(t).
\end{align}
Here, $\hat{x}_\text{B}(t) = e^{\im H_\text{B} t} \hat{x}_\text{B} e^{-\im H_\text{B} t}$ and $\expval{(\cdot)}_\text{B} \equiv \frac{\tr_\text{B} [(\cdot) e^{-\beta H_\text{B}}]}{\tr_\text{B}[e^{-\beta H_\text{B}}]}$ denotes the thermal averages over the bath with $\beta=1/(k_\text{B} T)$. We then have the fluctuation--dissipation theorem, \cite{Yan2005-annurev-physchem}
\begin{equation}\label{eqn:bath_correlation_function}
    \expval{\hat{x}_\text{B}(t)\hat{x}_\text{B}(0)}_\text{B} = \frac{1}{\pi} 
    \int_{-\infty}^{\infty} \dd{\omega} \frac{e^{-i \omega t}J(\omega)}{1 - e^{-\beta\omega}}.
\end{equation}




In this work, we use the {two-level system (TLS)} Hamiltonian in Eq.~\ref{eqn:H} to study a simple yet general relaxation process---an out-of-equilibrium spin one-half particle relaxes to equilibrium through the coupling to lattice vibrations. This thermal relaxation process depends on the system operator $\hat Q_\text{s}$, collective mode $\hat{x}_\text{B}$, and the bath polynomial. When $\hat Q_\text{s} \propto \sigma_z$, the TLS becomes the so-called pure decoherence model \cite{RevModPhys-Breuer2016}, representing the decoherence contribution from the coupling to lattice vibrations; while $\hat Q_\text{s} \propto \sigma_x \text{ or } \sigma_y$ denotes the spin relaxation contribution. Without loss of generality, we will restrict to the case of $\hat Q_\text{s} \propto \sigma_x$ to describe spin-lattice relaxation.

\subsection{\label{subsec:FGR} Fermi's golden rule rates}
In the following, we calculate the thermally averaged FGR rates for the TLS. The FGR rates for the transition from spin state $i$ to $f$ ($i, f = 0, 1$) are given by \cite{nitzan2013chemical}
\begin{equation}\label{eqn:FGR-discrete}
    k_{f \gets i}(\beta) = \frac{2\pi}{\hbar} \sum_a P_a(\beta) \sum_b \abs{W_{f,b \gets i,a}}^2 \delta(E_{f, b} - E_{i, a}).
\end{equation}
Here $i,a$ and $f,b$ denote coupled spin-vibrational states $\ket{i, a}$ and $\ket{f, b}$; $P_a(\beta) = \exp(-\beta \omega_a) / Z$ denotes the Boltzmann weight for the $a$-th vibrational state where $Z=\sum_a \exp(-\beta \omega_a)$ is the bare--bath partition function; and $W_{f, b \gets i, a} = \mel{f, b}{H_1}{i, a}$ denotes the transition matrix element. For the FGR calculation, we only consider $\hat{Q}_\text{s} = \sigma_x$, i.e., the relaxation contribution from the spin-lattice interaction. Thus, only two terms are non-vanishing: $W_{f, 1 \gets i, 0}=\mel{1, b}{z(\hat{x}_\text{B})}{0, a}$ and $W_{f, 0 \gets i, 1}=\mel{0, b}{z(\hat{x}_\text{B})}{1, a}$.  

The discrete Eq.~\ref{eqn:FGR-discrete} for $k_{f \gets i}(T)$ can be recast in terms of correlation function (see Eq.~\ref{eqn:discrete_to_continuous}):
\begin{equation}\label{eqn:FGR-continuous}
    k_{f \gets i} (\beta) = \int_{-\infty}^{\infty} \dd{t} e^{\pm \im \epsilon t} \expval{z(\hat{x}_\text{B}(t))z(\hat{x}_\text{B}(0))}_\text{B},
\end{equation}
where we use the upper sign ($+$ in this equation) in the exponential denotes the $1\gets0$ process, and the lower sign ($-$) denotes the $0\gets1$ process, respectively. The upper and lower sign convention will be consistently applied henceforth. Here, $\epsilon$ denotes the energy difference between states 0 and 1. In other words, the FGR rates depend on the time correlation functions of polynomial $z(\hat{x}_\text{B})$. 
In moderate coupling strength, we are safe to truncate the polynomial up to the second order that $z(x) = \alpha_0 + \alpha_1 x + \alpha_2 x^2$. Additionally, we can also absorb the zero-th order interaction $\alpha_0$ into the quantum system $H_\text{s}$. When $\alpha_0=0$, the spin states $\ket{0}$ and $\ket{1}$ are eigen states, hence $\epsilon = \Delta$. Otherwise we have to calculate $\epsilon$ from diagonalize $H_s + \alpha_0 \sigma_x$. With these considerations, the calculations of $T^{-1}$ using FGR becomes straightforward, which are outlined in Appendix~\ref{app:goldenrule}. It is noteworthy that two terms,
\[ 
    \alpha_1^2 \expval{\hat{x}_\text{B}(t) \hat{x}_\text{B}(0)}_\text{B}, \text{ and }
    \alpha_2^2 \expval{\hat{x}_\text{B}^2(t) \hat{x}_\text{B}^2(0)}_\text{B}
,\]
prove to be finite, while the cross linear-quadratic terms in the correlation function vanish. After some algebra, we obtain closed expressions for the FGR relaxation rates given by 
\begin{subequations}\label{eqn:FGR_results}
\begin{gather}
    k_{f \gets i} = \alpha_1^2 k_{f \gets i}^{\text{(1)}} + \alpha_2^2 k_{f \gets i}^{\text{(2)}}, \\
    k_{f \gets i}^{\text{(1)}} = 2\frac{J(\pm \epsilon)}{1 - e^{\mp \beta \epsilon}}, 
    \\
    k_{f \gets i}^{\text{(2)}} = \frac{4}{\pi} \int_{-\infty}^{\infty} \dd{\omega}
    \frac{J(\omega)}{1 - e^{-\beta \omega}} \frac{J(\omega \pm \epsilon)}{e^{\beta (\omega \pm \epsilon)} - 1}, 
\end{gather}
\end{subequations}
with superscript $(1)$ and $(2)$ denoting $\hat{x}_\text{B}$ and $\hat{x}_\text{B}^2$ coupling term \cite{Lunghi2023}, respectively.

Notably, {an oscillating term appears from the quadratic contribution to the rates}
\begin{equation}\label{eqn:oscillatory_term}
{
    \left[\int_{0}^{\infty} \frac{\dd{\omega}}{\pi} J(\omega) \coth{\frac{\beta\omega}{2}}\right]^2 \int_{-\infty}^{\infty} \dd{t} e^{\pm \im \epsilon t},
}
\end{equation}
{which is not incorporated to the rate equations (Eqs.~\ref{eqn:FGR_results}). This is because the second oscillating integral in } this term {does not converge} to a finite value. Intriguingly, we do observe oscillating behaviors of the rate kernels in the extended DEOM simulations, as discussed in Sec.~\ref{sec:results}.

\subsection{\label{subsec:quad_environment} The extended DEOM for both linear and quadratic environments}
In the standard HEOM formalism, only the linear coupling to the bath mode $\hat{x}_\text{B}$ is considered.\cite{EDEOM2-Xu2018, Perspective-Tanimura2020} However, to accurately model the spin-lattice interaction dynamics requires the explicit considerations of non-linear coupling at least to the second order ($\hat{x}_\text{B}^{2}$), since 
the relaxations with two phonon processes are very common in experiments. Fortunately, the extended dissipaton equation of motion (DEOM) formalism, by R.-X. Xu and colleagues, extends the HEOM methods for non-linear interactions up to the quadratic order.\cite{EDEOM1-Xu2017, EDEOM2-Xu2018, Core-System-Chen2023} This study employs the extended DEOM formalism to investigate the spin-lattice relaxation. Subsequently, we provide a concise introduction to this method.

Similar to the HEOM formalism, the extended DEOM requires to expand the bath correlation function in finite exponential series, 
\begin{equation}\label{eqn:exp_series}
    \expval{\hat{x}_\text{B}(t)\hat{x}_\text{B}(0)}_\text{B} = \sum_k^{K} \eta_k e^{-\gamma_k t},
\end{equation}
where $K$ sets of parameters $\eta_k$ and $\gamma_k$ are generally obtained from fitting or spectral function decomposition such as the Pad\'{e} decomposition. \cite{PADE-Hu2010, Prony-Chen2022} In contrast to the auxiliary density matrix interpretation of HEOM, DEOM employs a quasi-particle interpretation of such decomposition that leads to the following algebra 
\begin{gather}
    \hat{x}_\text{B} = \sum_k^{K} \hat{f}_k, \\
    \expval{\hat{f}_k(t) \hat{f}_{k'}(0)}_\text{B} = \delta_{kk'} \eta_k e^{-\gamma_k t}.
\end{gather}
Here, $\{\hat{f}_k\}_{k=1}^K$ are the so-called dissipaton operators, where 
the dissipatons are independent quasi-particles that represent the collective bath.

With the dissipaton operators, we can represent the total system dynamics as a set of equations of motion for the dissipaton density operators (DDOs). To begin, we know the composite system density operator $\rho_\text{T}$ satisfies the von Neumann equation
\begin{equation}
    \dot{\rho}_\text{T}(t) = - \im  \comm{H_\text{T}}{\rho_\text{T}(t)}.
\end{equation}
The DDOs can be defined as
\begin{equation}
    \rho_{\bm{n}}^{(n)}(t) = \rho_{n_1 \dots n_K}^{(n)}(t) = \tr_\text{B}[(\hat{f}_K^{n_K} \dots \hat{f}_1^{n_1})^{\circ} \rho_\text{T}(t)].
\end{equation}
Here, index $\bm{n} \equiv\{ n_1, \dots,  n_K\}$ denotes the configuration, and $n = n_1 + \dots + n_K$ denotes the total number of disspaton exitations. 
The notation $(\cdots)^{\circ}$ denotes the irreducible representation. For bosonic dissipaton, as is in the case of lattice vibration, we have $(\hat{f}_k \hat{f}_{k'})^{\circ} = (\hat{f}_{k'} \hat{f}_{k})^{\circ}$.
With these, we can know the reduced system density operator are given by $\rho_\text{S} \equiv \rho_{\bm{0}}^{(0)}$.
Refs.~\cite{EDEOM1-Xu2017, EDEOM2-Xu2018} show that the reduced system dynamics given by Eq.~\ref{eqn:H1} is obtained by applying the dissipaton algebra, including the generalized generalized Wick's theorems (GWT-I and GWT-II). The equations of motion of $\{\rho_{\bm n}^{(n)}\}$ read
\begin{equation}
\label{eqn:extended_DEOM_EOMs}
\begin{aligned}
    \dot{\rho}_{\bm{n}}^{(n)} &= 
    -(\im \mathcal{L}_\text{s} + \sum_k n_k \gamma_k)\rho_{\bm{n}}^{(n)}
    -\im (\alpha_0 + \alpha_2 \expval{\hat{x}_\text{B}^2}_\text{B}) \mathcal{A} \rho_{\bm{n}}^{(n)}  \\ 
    & -\im\alpha_1\sum_k(\mathcal{A} \rho_{\bm{n}_k^{+}}^{(n+1)} + n_k\mathcal{C}_k \rho_{\bm{n}_k^{-}}^{(n-1)}) 
    -2\im\alpha_2\sum_{kk'} n_k \mathcal{C}_k \rho_{\bm{n}_{kk'}^{+-}}^{(n)} \\
    & -\im\alpha_2\sum_{kk'} \left[\mathcal{A} \rho_{\bm{n}_{kk'}^{++}}^{(n-2)} + n_k (n_{k'} - \delta_{kk'}) \mathcal{B}_{kk'} \rho_{\bm{n}_{kk'}^{--}}^{(n-2)} \right],
\end{aligned}
\end{equation}
with some of the convenient operators used above defined as the following,
\begin{subequations}
\begin{gather}
    \mathcal{L}_\text{s} \hat{O} \equiv \comm{H_\text{s}}{\hat{O}}, \\
    \mathcal{A} \hat{O} \equiv \comm{\hat{Q}_\text{s}}{\hat{O}}, \\
    \mathcal{B}_{kk'} \hat{O} \equiv 
    \eta_k \eta_{k'} \hat{Q}_\text{s} \hat{O} - \eta_{\bar{k}}^{*} \eta_{\bar{k}'}^{*} \hat{O} \hat{Q}_\text{s}, \\
    \mathcal{C}_{k} \hat{O} \equiv 
    \eta_k \hat{Q}_\text{s}\hat{O} - \eta_{\bar{k}}^{*}\hat{O}\hat{Q}_\text{s}.
\end{gather}
\end{subequations}

For later use, we introduce a compact notation, 
\begin{equation}\label{eqn:compactDEOM}
    \dot{\bm{\rho}}(t) = - \im \calbm{L}(t) \bm{\rho}(t).
\end{equation}
Here, $\bm{\rho}(t) = \{\rho_{\bm{n}}^{(n)}; n = 0, 1, 2, \dots \}$ denotes the set of DDOs. The equations of motion encoded in Eq.~\ref{eqn:extended_DEOM_EOMs} are represented by the DEOM-space dynamics generator $\calbm{L}(t)$. 

\subsection{\label{subsec:rate_kernel} Rate kernel from the extended DEOM formalism}
As introduced in Sec.~\ref{subsec:quad_environment}, DDO equations of motions (Eq.~{\ref{eqn:extended_DEOM_EOMs}}) provide a numerically exact recipe for the reduced system dynamics. We can go one step further to extract the rate kernels for the spin-lattice relaxation by applications of the projection operator techniques. \cite{rate-Zhang2016} 

Following ~ref. \cite{rate-Zhang2016}, we use the Nakajima-Zwanzig projection operator technique onto the DDOs. \cite{Projection-Nakajima1958, Projection-Zwanzig1960}
For the reduced system density $\rho^{(0)}$, it's intuitive to define $\mathcal{P}$ and $\mathcal{Q} \equiv \one - \mathcal{P}$ as the following
\begin{equation}
    \begin{gathered}
        \mathcal{P} \rho^{(0)} = P_0(t) \ketbra{0}{0} +  P_1(t) \ketbra{1}{1}, \\
        \mathcal{Q} \rho^{(0)} = \rho^{(0)}_{01}(t) \ketbra{0}{1} + \rho^{(0)}_{10}(t) \ketbra{1}{0},
    \end{gathered}
\end{equation}
where $\mathcal{P}$ and $\mathcal{Q}$ represents the population and coherence, respectively. The projection definitions can then be generalized to the full DDO-space by
\begin{equation}
\begin{gathered}
	\calbm{P} \bm{\rho}(t)	= \{\mathcal{P} \rho^{0}(t); 0, 0, \dots\} \equiv \bm{p}(t), \\
	\calbm{Q} \bm{\rho}(t)	= \{\mathcal{Q} \rho^{0}(t); \rho^{(n>0)}_{\bm{n}} (t) \} \equiv \bm{\sigma}(t),  
\end{gathered}
\end{equation}
such that Eq.~\ref{eqn:compactDEOM} can be recasted into:
\begin{equation}
    \label{eqn:HEOM-P}
    \begin{pmatrix}
		\dot{\bm{p}} \\
		\dot{\bm{\sigma}}
	\end{pmatrix} = -\im%
	\begin{pmatrix}
		\calbm{PLP} & \calbm{PLQ} \\
		\calbm{QLP} & \calbm{QLQ}
	\end{pmatrix}%
	\begin{pmatrix}
		\bm{p} \\
		\bm{\sigma}
	\end{pmatrix}.
\end{equation}
Formally, one can integrate the equation for the $\calbm{Q}$-space density $\bm{\sigma}$ and obtain solution:
\begin{equation}
    \bm{\sigma}(t) = e^{-\im \calbm{QL}t} \bm{\sigma}(0) - %
    \im \int_0^t \dd{\tau} e^{-\im \calbm{QL}(t - \tau)} \calbm{QL} \bm{p}(\tau).
\end{equation}
By inserting this formal solution of $\bm{\sigma}(t)$ to the equations of motion for $\calbm{P}$-space density $\bm{p}$, one obtains:
\begin{equation}
\label{eqn:kernel}
\begin{gathered}
    \dot{\bm{p}}(t) = -\im \calbm{PL} e^{-\im \calbm{QL}t} \bm{\sigma}(0) +%
        \int_0^t \dd{\tau} \calbm{K}(t-\tau) \bm{p}(\tau), \\
    \calbm{K}(t) = -\calbm{PL}e^{-i\calbm{QL}t} \calbm{QLP}.
\end{gathered}
\end{equation}
In practice, the standard factorization $\bm{\rho}(0) = \rho_\text{S}(0) \otimes \rho_\text{B}^\text{eq}$ of the initial condition is used, meaning $\rho^{(n>0)}_{\bm{n}} (t) = 0$. 
As a result, projected system density $\bm{p}(t)$ satisfies intergo-differential equation
\begin{equation}\label{eqn:compact_kernel}
    \dot{\bm{p}}(t) = \int_0^t \dd{\tau} \calbm{K}(t-\tau) \bm{p}(\tau).
\end{equation}

The rate kernel $\calbm{K}(t)$ encodes the full non-Markovian dynamics of the reduced system, and we can readily extract the $T_1^{-1}$ rate from $\calbm{K}(t)$ . In particular, the reduced population of spin state $a$ ($a=0, 1$) satisfies the following kinetic equation:
\begin{equation}
\begin{aligned}\label{eqn:kernel_component_form}
    \dot{P}_a(t) = - & \int_0^t \dd{\tau} \mathcal{K}_{b \gets a}(t-\tau) P_a(\tau) + \\
                     & \int_0^t \dd{\tau} \mathcal{K}_{a \gets b}(t-\tau) P_b(\tau) .
\end{aligned}
\end{equation}
In the Markovian limit, the above kinetic equation becomes simpler: $\dot{P}_1(t) = k_{0 \gets 1} P_1(t) + k_{1 \gets 0} P_0(t)$, where 
\begin{equation}
    k_{a \gets b} = \int_{0}^{\infty} \dd{\tau} \mathcal{K}_{a \gets b}(\tau), \quad a, b \in \{0, 1\}.
\end{equation}
Here, $k_{0\gets1}$ and $k_{1\gets0}$ encode the long time behaviour of the rate kernel components $\mathcal{K}_{0\gets1}$ and $\mathcal{K}_{1\gets0}$. 
In the context of spin-lattice relaxation, 
\begin{equation}\label{eqn:T1_from_mark}
    T_1^{-1} = k_{0\gets1} + k_{1\gets0}.
\end{equation}

\section{\label{sec:results} Results and Discussions}

\begin{figure*}[htbp]
    \centering
    \includegraphics[scale=.75]{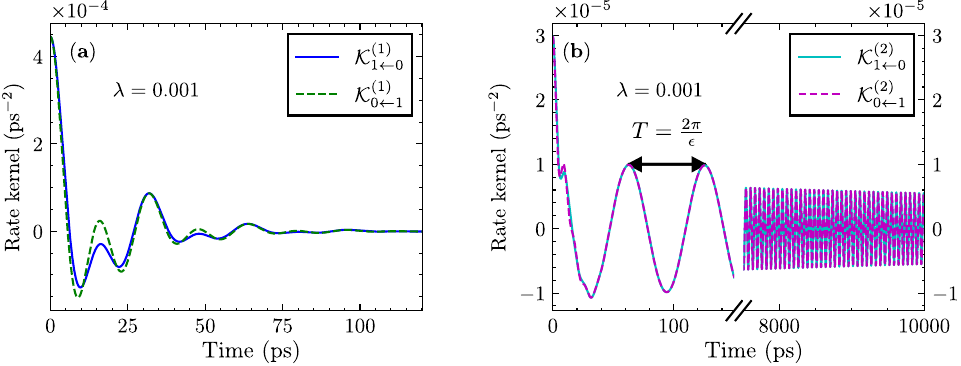}
    \includegraphics[scale=.75]{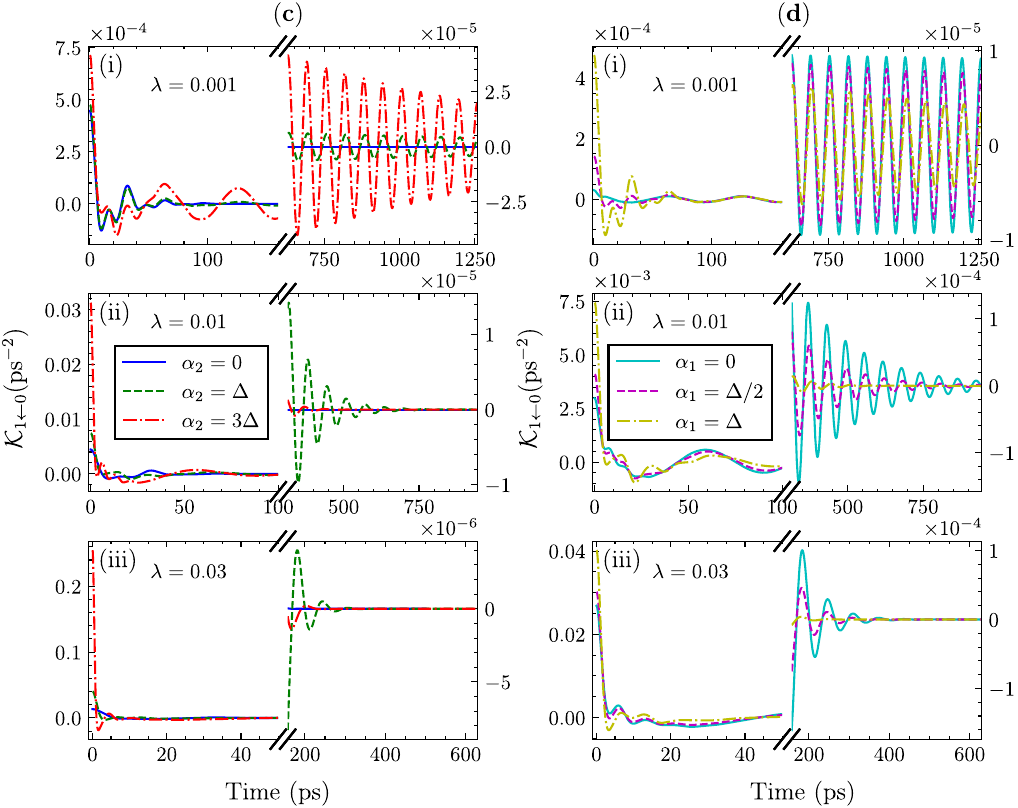}
    \caption{
Characteristic rate kernels illustrating $\hat{x}_\text{B}$-, $\hat{x}_\text{B}^2$-, and $\alpha_1 \hat{x}_\text{B} + \alpha_2 \hat{x}_\text{B}^2$ spin-lattice interactions. Panel (a) displays the kernel with exclusive $\hat{x}_\text{B}$-coupling ($\alpha_1 = \unitval{10/3}{\wavenumber}$, $\alpha_2 = \unitval{0}{\wavenumber}$). Panel (b) illustrates the scenario of exclusive $\hat{x}_\text{B}^2$-coupling ($\alpha_1 = \unitval{0}{\wavenumber}$, $\alpha_2 = \unitval{10/3}{\wavenumber}$). Notably, both $\mathcal{K}_{1\gets0}^{(2)}$ and $\mathcal{K}_{0\gets1}^{(2)}$ initiate free induction decay like profile with a frequency $\epsilon = \Delta$ after initial dynamics. In panels (a) and (b), $\lambda = 0.001$. Panel (c) portrays the impact of $\alpha_2$ on $\mathcal{K}^{(1)}$ while keeping $\alpha_1 = \unitval{10/3}{\wavenumber}$. Panel (d) illustrates the influence of $\alpha_1$ on $\mathcal{K}^{(2)}$ with $\alpha_2 = \unitval{10/3}{\wavenumber}$. In panels (c) and (d), sub-panels (i-iii) correspond to $\lambda=0.001, 0.01, 0.03$, respectively. Only the $0\to1$ rate is plotted for clarity. The following parameters are consistent across all four panels: $\alpha_0 = \unitval{0}{\wavenumber}$, $\Delta = \zeta= \unitval{10/3}{\wavenumber}$, $\omega_\text{B}=\unitval{10}{\wavenumber}$, and temperature $T=\unitval{53.2}{\kelvin}$.
    }
    \label{fig:kernel-vs-time}
\end{figure*}

In the following, we calculate the rate kernel for the spin-lattice relaxation dynamics using the extended DEOM formalism described in Sec.~\ref{subsec:quad_environment} and \ref{subsec:rate_kernel}. Here we consider then rate kernel of solely linear (denoted as $\mathcal{K}^{(1)}$), solely quadratic ($\mathcal{K}^{(2)}$) and general mixed linear-quadratic ($\mathcal{K}$) system-bath coupling, respectively. With the rate kernel results, we evaluate the Markovian limit of the $T_1^{-1}$ rates given by Eq.~\ref{eqn:T1_from_mark}, and make a comparison with the FGR rates given by Eqs~\ref{eqn:FGR_results}. 

{Throughout this section, we employ a Brownian motion spectral function to describe the collective lattice mode. In particular, this model can describe} the spin-relaxation through a local lattice vibrational mode with a characteristic frequency of $\omega_\text{B}$, meanwhile this characteristic mode couples to the bath modes. This type of system-oscillator-bath is widely applied to study dissipative dynamics, \cite{TwoLevelDissipative-Leggett1987, BOspectral-Garg1985, BO-Tanaka2009} and can be readily mapped to the simple spin-boson model (Eq.~\ref{eqn:H}) via the well-established canonical transformation \cite{Leggett1984, BOspectral-Garg1985}. To this end, the local mode relaxation scheme can be described by a Brownian motion spectral function \cite{EDEOM1-Xu2017} given by
\begin{equation}\label{eqn:BO_spectra}
    J(\omega) = \Im \frac{2 \lambda \omega_{B}^{2}}{\omega_{B}^{2} - \omega^{2} - i \omega \zeta}.
\end{equation}
Here, $\lambda$ denotes the interaction strength, and we assume the friction function $\zeta(\omega) = \zeta$ is frequency independent \cite{BO-Tanaka2009, Core-System-Chen2023}. With expression Eq.~\ref{eqn:BO_spectra} for $J(\omega)$, we construct the extended DEOM by decomposing correlation function Eq.~\ref{eqn:bath_correlation_function} into a few dissipatons using the time-domain Prony fitting method. \cite{Prony-Chen2022}

\subsection{\label{subsec:kernel-linear-vs-quad} Rate kernels of linear and quadratic coupling}

In FIG.~\ref{fig:kernel-vs-time}, we present characteristic rate kernels for the spin-lattice relaxation dynamics. Specifically, we consider a spin with Zeeman energy $\Delta = \unitval{10/3}{\wavenumber}$, of the characteristic energy of W-band ($100 \mathrm{GHz}$) Electron Paramagnetic Resonance (EPR) experiment. This spin couples to a characteristic local phonon mode centered at frequency $\omega_\text{B}=\unitval{10}{\wavenumber}$, {corresponding to a spin-lattice dissipation channel in the terahertz regime.}

Notably, the behaviors of the rate kernel as a function of time are drastically different for $\hat{x}_\text{B}$- and $\hat{x}_\text{B}^{2}$- spin-lattice coupling. The linear kernel $\mathcal{K}^{(1)}$ completely decays (FIG.~\ref{fig:kernel-vs-time} (a)) within a few picoseconds, which is the typical time it takes for lattice phonon oscillation. In contrast,
the quadratic kernels $\mathcal{K}^{(2)}$ continue to oscillate without much decay (FIG.~\ref{fig:kernel-vs-time} (b)), when the coupling is weak. Moreover, the long time limit of the $\mathcal{K}^{(2)}$ kernels exhibits a free induction decay (FID) like feature, with a frequency $\epsilon = \Delta$, despite the initial dynamics have a more complicated structure.

As we indicated in Sec.~\ref{subsec:FGR}, the FGR rates for the $\hat{x}_\text{B}^2$ coupling indeed contain an oscillating term (Eq.~\ref{eqn:oscillatory_term}), which agrees with the FID-like feature of the $\mathcal{K}^{(2)}$ kernels calculated from extended HEOM. However, the FGR calculations do not capture the decay of the oscillating term, so the oscillating terms cannot convert to a finite value and are thus not included in the final rate calculation. This indicates simple perturbation theories ignore some contributions to the rates. In particular, these ignored interactions can somehow damp the oscillating integral Eq.~\ref{eqn:oscillatory_term} term. To understand the effect of the neglected contribution, we propose a simple model in Appendix.~\ref{app:demo_importance_of_decaying}. The simple model demonstrates when the decay rate, denoted  $\Gamma$, is comparable with the oscillation frequency, the neglected contribution can significantly alter the relaxation rate.

The extended DEOM simulations reveal that the decay rate $\Gamma$ of the FID-like kernel $\mathcal{K}^{(2)}$ depends on the coupling strength $\lambda$, and can be notably enhanced when the quadratic coupling is mixed with linear coupling. As illustrated in FIG.~\ref{fig:kernel-vs-time} (c.i-iii), the decay rate $\Gamma$ increases with the coupling strength. In addition, panel (d) demonstrates when linear coupling is added to quadratic coupling, the decay rate $\Gamma$ is significantly enhanced when compared with the purely $\hat{x}_\text{B}^2$-coupling (panel (b)). Notably, the decay is greatly enhanced even for weak coupling ($\lambda = 0.001$). Overall, these comparisons indicate the FGR rates ignores 1) all the higher-order terms in the perturbation series, and 2) the interaction between the linear and quadratic coupling term. On the other hand, the numerical exact DEOM simulations capture the higher order terms in the perturbation series and the interaction between the linear and quadratic coupling term (Eq.~\ref{eqn:extended_DEOM_EOMs}).

Overall, we demonstrate a stronger coupling and a mixed linear quadratic interaction can enhance the relaxation rate of the FID-like rate kernel. As we soon demonstrate in Section~\ref{subsec:rate_temperature}, this subtlety becomes a source of contribution that leads to the failure of FGR rate estimations, which can qualitatively alter the temperature dependence of $T_1^{-1}$.

\subsection{\label{subsec:rate_temperature} The temperature dependency of $T^{-1}$ rate}
With rate kernels readily obtained by the extended DEOM formalism, we are in the position to evaluate the $T_1^{-1}$ rate using Eq.~\ref{eqn:T1_from_mark}. In particular, we focus on how $T_1^{-1}$ depends on temperature.

\begin{figure}[htbp]
    \centering
    \includegraphics[scale=0.7]{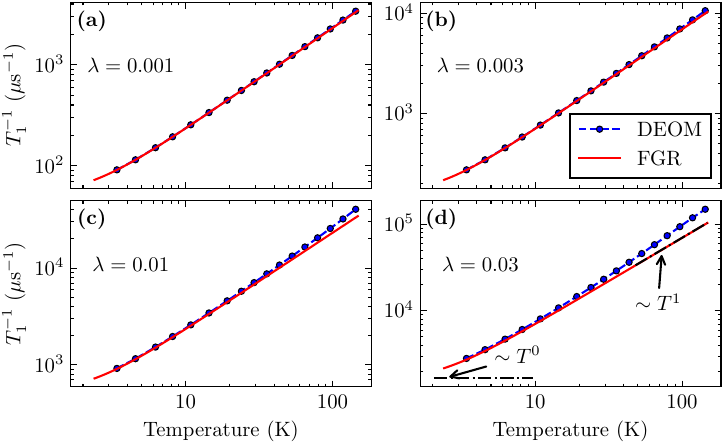}
    \caption{
$T_1^{-1}$ as a function of temperature in solely $\hat{x}_\text{B}$-coupling. Panels (a-d) correspond to interaction strength $\lambda = 0.001, 0.003, 0.01, 0.03$, respectively. The following parameters are used: $\alpha_1 = \unitval{10/3}{\wavenumber}$, $\alpha_0 = \alpha_2 = \unitval{0}{\wavenumber}$, $\Delta = \zeta= \unitval{10/3}{\wavenumber}$, $\omega_\text{B}=\unitval{10}{\wavenumber}$.
    }
    \label{fig:linear-T1-rate}
\end{figure}

\begin{figure}[htbp]
    \centering
    \includegraphics[scale=0.7]{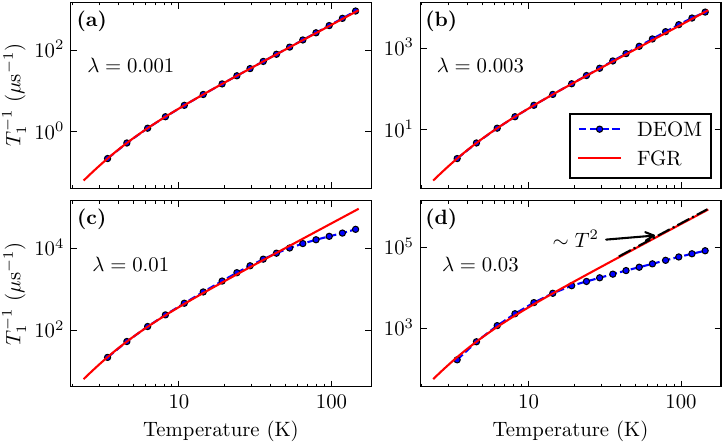}
    \caption{
$T_1^{-1}$ as a function of temperature in solely $\hat{x}_\text{B}^2$-coupling. Panels (a-d) correspond to interaction strength $\lambda = 0.001, 0.003, 0.01, 0.03$, respectively. The following parameters are used: $\alpha_2 = \unitval{10/3}{\wavenumber}$, $\alpha_0 = \alpha_1 = \unitval{0}{\wavenumber}$, $\Delta = \zeta= \unitval{10/3}{\wavenumber}$, $\omega_\text{B}=\unitval{10}{\wavenumber}$.
    }
    \label{fig:quadratic-T1-rate}
\end{figure}

FIGs.~\ref{fig:linear-T1-rate} and~\ref{fig:quadratic-T1-rate} compare $T_1^{-1}$ predicted by the FGR and DEOM for the solely linear and solely quadratic coupling case, respectively. Notably, the FGR results agree with that of DEOM in the weak coupling limits across a wide range of temperatures in both cases, despite the quadratic dynamics clearly show long term memories (FIG.~\ref{fig:kernel-vs-time}). Such agreement indicates when the interaction strength $\lambda$ is small, the higher order terms in the perturbation series, and the damping of the oscillating rate terms are not important. Consequently, in the weak coupling limit, the temperature scaling of $T_1^{-1}$ for linear and quadratic coupling are $T$ and $T^2$, respectively, to which both the FGR and DEOM results agree.

Nevertheless, FGR rates indeed deviate from DEOM when stronger coupling is present. In particular, panel (c) and (d) of FIGs.~\ref{fig:linear-T1-rate} and~\ref{fig:quadratic-T1-rate} indicate FGR tend to underestimate the $T_1^{-1}$ in linear coupling, and overestimate the $T_1^{-1}$. These panels demonstrate the well known high temperature scaling laws fails when the coupling is strong. Overall, the quadratic rates tend to have stronger deviations than the linear rates.

\begin{figure}[htbp]
    \centering
    \includegraphics[scale=0.7]{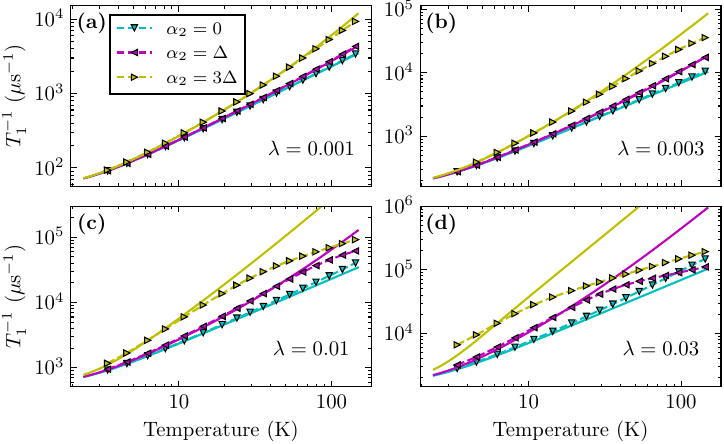}
    \caption{
        The effect of $\alpha_2$ on $T_1^{-1}$ as a function of temperature when $\alpha_1=\unitval{10/3}{\wavenumber}$ is fixed. Panels (a-d) correspond to interaction strength $\lambda = 0.001, 0.003, 0.01, 0.03$, respectively. Other parameters are: $\alpha_0 = \unitval{0}{\wavenumber}$, $\Delta=\zeta=\unitval{10/3}{\wavenumber}$, $\omega_\text{B}=\unitval{10}{\wavenumber}$. 
    }
    \label{fig:linear-mixed-quad}
\end{figure}

\begin{figure}[htbp]
    \centering
    \includegraphics[scale=0.7]{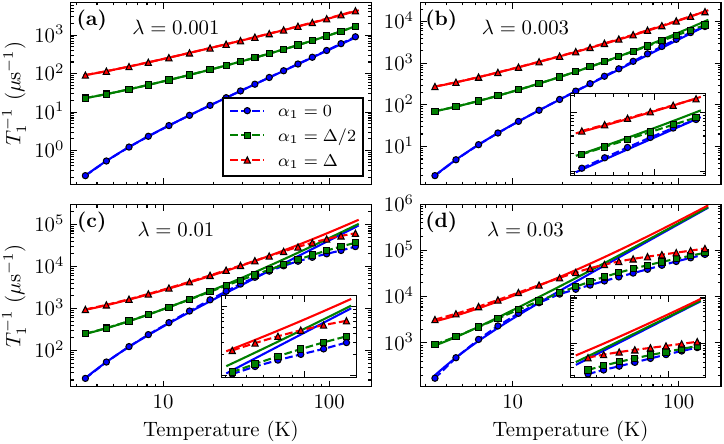}
    \caption{
The effect of $\alpha_1$ on $T_1^{-1}$ as a function of temperature when $\alpha_2=\unitval{10/3}{\wavenumber}$ is fixed. Panels (a-d) correspond to interaction strength $\lambda = 0.001, 0.003, 0.01, 0.03$, respectively. Other parameters are: $\alpha_0 = \unitval{0}{\wavenumber}$, $\Delta=\zeta=\unitval{10/3}{\wavenumber}$,$\omega_\text{B}=\unitval{10}{\wavenumber}$.}
    \label{fig:quad-mixed-linear}
\end{figure}

Finally, we demonstrate that more dramatic deviations can occur when we consider the case of a mixed linear quadratic spin-lattice relaxation channel. In particular, FIG.~\ref{fig:linear-mixed-quad} demonstrates the FGR overestimates both the values and scaling of $T_1^{-1}$ at high-temperature regime, when we add some quadratic coupling when $\alpha_1$ is fixed. In comparison, when $\alpha_2$ is fixed and some linear character is added to the spin-lattice interaction, we observe a lesser degree of deviation. (FIG.~\ref{fig:quad-mixed-linear})

Overall, we demonstrate when the spin-lattice coupling is strong, we cannot simply treat the dynamics and rates with perturbation theory. In addition, although the linear and quadratic coupling are generally treated as independent in literature, \cite{Lunghi2023} we demonstrate the interactions are very important for the $T_1^{-1}$ rates when the coupling is somewhat stronger.

\section{\label{sec:conclusion} Conclusion}
In this work, we apply the extend DEOM method to examine the validity of FGR in modeling the dynamics of spin-lattice interaction. The numerical exact DEOM approach reveals the free induction decay (FID) feature of quadratic coupling dynamics. The decay rate of the kernel can be {rather slow [FIG.~\ref{fig:kernel-vs-time} (b)]}, indicating strong non-Markovian nature for the two-phonon processes encoded in $\hat{x}_\text{B}^2$-coupling. We demonstrate this damping rate depends on 1) the coupling strength, and 2) the interactions between the one and two phonon processes. Indeed, damping of the rate kernel is completely neglected in perturbation treatments as they are encoded in higer order interactions. Consequently, methods such as FGR and Markovian master equations fail to correctly predict the relaxation dynamics and the temperatures dependencies when the coupling is strong.

Looking forward, these findings as well as the methods presented in this work should be very useful in {understanding} qubits relaxation dynamics in the field of quantum information. One particular applications of {our methodology is the study of spin-lattice relaxation in the NV center},\cite{NV-Norambuena2018}  is a three level system that can encode more complicated and interesting dynamics. {
Furthermore, systems with large spin-orbit couplings exhibit strong spin-lattice interaction, \cite{strong-couping-Grigoryev2021, demonstrate_strong-Xu2024, rev-strong-Lu2024} it will be interesting to see whether strong couplings play an important role for these materials.} In addition, the method here can be coupling with data driven methods such as DMD,\cite{DMD-Liu2023} provide the possibility of studying spin-relaxation dynamics for systems of large dimension. 
\vspace{5em}

\begin{acknowledgments}
W.D. thanks the funding from National Natural Science Foundation of China (No. 22361142829) and Zhejiang Provincial Natural Science Foundation (No. XHD24B0301). L.S. thanks supports from the National Natural Science Foundation of China (No. 22273078) and Hangzhou Municipal Funding, Team of Innovation (No. TD2022004). Y.W thanks the support from the National Natural Science Foundation of China (Nos. 22103073 and 22373091). We thank Westlake university supercomputer center for the facility support and technical assistance.
\end{acknowledgments}

\appendix

\section{\label{app:goldenrule} Fermi's golden rule calculation of $T_1$}
The discrete FGR rate (Eq.~\ref{eqn:FGR-discrete}) is recast into integration of continuous correlation function:
\begin{widetext}
\begin{align} \label{eqn:discrete_to_continuous}
    k_{f \gets i} (\beta)
    &= \int_{-\infty}^{\infty} \dd{t} \sum_{a, b} \frac{e^{-\beta \omega_a}}{Z}  \mel{a}{z(\hat{x}_\text{B})}{b} \mel{b}{z(\hat{x}_\text{B})}{a} e^{\im (E_b - E_a \mp \epsilon) t},\\
    &= \int_{-\infty}^{\infty} \dd{t} \frac{e^{\pm \im \epsilon t}}{Z} \sum_{a} \mel{a}{e^{\im H_\text{B} t} z(\hat{x}_\text{B}) e^{-\im H_\text{B} t} z(\hat{x}_\text{B}) e^{-\beta H_\text{B}}}{a}, \\
    &= \int_{-\infty}^{\infty} \dd{t} e^{\pm \im \epsilon t} \expval{z(\hat{x}_\text{B}(t))z(\hat{x}_\text{B}(0)}_\text{B}.
\end{align}
\end{widetext}
where $\pm$ signs denote the $1\gets0$ and $0\gets1$ process, respectively. Here, $\epsilon$ denotes energy difference between state 0 and 1. When $\alpha_0 = 0$, $\epsilon = \Delta$. Otherwise we diagonalize $H_\text{s'} = \frac{\Delta}{2} \sigma_z + \alpha_0 \sigma_x$ to obtain $\epsilon$.

In this work, we truncate polynomial to $z(x) = \alpha_1 x + \alpha_2 x^2$. We now show that many terms in the correlation function vanish. Particularly, terms with odd number of bath operators vanish, e.g., three operator terms like $\expval{\hat{x}_\text{B}^2(t) \hat{x}_\text{B}(0)}_\text{B}$. This is because when taking thermal averages of non-interacting boson operators, three term creation/annihilation operators averages vanish. To demonstrate this, we suppose $\hat{x}_\text{B} \equiv \sum_k c_k \hat{q}_k$ and $\hat{q}_k = \frac{1}{\sqrt{2}}(\hat{b}^{\dagger}_k + \hat{b}_k)$. For example, correlation function $\expval{\hat{x}_\text{B}^2(t) \hat{x}_\text{B}(0)}_\text{B}$ will consist of three operator term such as
\begin{equation*}
    \expval{\hat{b}_j \hat{b}_k \hat{b}_l^{\dagger}}_\text{B}  
    \propto \Tr[\hat{b}_j \hat{b}_k \hat{b}_l^{\dagger} e^{-\beta H_\text{B}}].
\end{equation*}
Using identities $\hat{b}_l^{\dagger} e^{-\beta H_\text{B}} = e^{-\beta H_\text{B}} e^{\beta H_\text{B}} \hat{b}_l^{\dagger} e^{-\beta H_\text{B}} = e^{\beta \omega_l} e^{-\beta H_\text{B}} \hat{b}_l^{\dagger}$, cyclic property of trace and commutation relations, we can can show these three-operator averages indeed vanishes. Hence, $k_{f \gets i}$ has only two contributions --- $\alpha_1^2 \expval{\hat{x}_\text{B}(t) \hat{x}_\text{B}(0)}_\text{B}$, and $\alpha_2^2 \expval{\hat{x}_\text{B}^2(t) \hat{x}_\text{B}^2(0)}_\text{B}$, which we refer to as the linear term ($k_{f \gets i}^{(1)}$) and quadratic term ($k_{f \gets i}^{(2)}$), respectively.

To calculate $k_{f \gets i}^{(1)}$, we just substitute Eq.~\ref{eqn:bath_correlation_function} into Eq.~\ref{eqn:FGR-continuous} and obtain
\begin{equation}\label{eqn:rate_linear}
k_{f \gets i}^{(1)}(\beta) = 
    2 \int_{-\infty}^{\infty} \dd{\omega} \frac{J(\omega)}{1 - e^{-\beta\omega}} \delta(\omega \pm \epsilon).
\end{equation}
Using the fact $J(\omega)$ is an odd function, we conclude the linear term contribution to $T_1^{-1}$ is $k_{1 \gets 0}^{(1)} + k_{0 \gets 1}^{(1)} = 2 J(\epsilon) \coth{(\frac{\beta \epsilon}{2})}$. It is easy to verify detailed balance $k_{0 \gets 1} = e^{-\beta \epsilon} k_{1 \gets 0}$ is satisfied for the FGR rates. Notably, we observe the temperature scaling of $k^{(0)}_{f\gets i}$ does not depend on the spectral density:
In the high temperature limit ($\beta \epsilon \ll 1$), $T_1^{-1} \propto T^{-1}$; In the low temperature limit ($\beta \to \infty$), $T_1^{-1}$ become temperature independent that $T_1^{-1} \propto T^{0}$.

To calculate $k_{f \gets i}^{(2)}$, we need first evaluate $\expval{\hat{x}_\text{B}^2(t) \hat{x}_\text{B}^2(0)}_\text{B}$. Specifically, using $\hat{x}_\text{B}(t) = \sum_j \frac{c_j}{\sqrt{2} } (\hat{b}_j^{\dagger} e^{\im \omega_j t} + \hat{b}_j e^{-\im \omega_j t})$, definition of the spectral function (Eq.~\ref{eqn:spectralfunction}), the fact that $J(\omega)$ is a odd function, and Wick's theorem, we obtain
\begin{widetext}    
\begin{equation}
\begin{aligned}\label{eqn:quad_correlation_function}
    \expval{\hat{x}_\text{B}^2(t) \hat{x}_\text{B}^2(0)}_\text{B} =
    &%
    2\int_{0}^{\infty} \frac{\dd{\omega}}{\pi} \frac{J(\omega)}{e^{\beta \omega} - 1}
    \int_{-\infty}^{\infty} \frac{\dd{\omega'}}{\pi} \frac{J(\omega')}{1 - e^{-\beta \omega'}} 
    \exp[\im(\omega - \omega')t] + \\
    &%
    2\int_{0}^{\infty} \frac{\dd{\omega}}{\pi} \frac{J(\omega)}{1 - e^{-\beta \omega}}
    \int_{-\infty}^{\infty} \frac{\dd{\omega'}}{\pi} \frac{J(\omega')}{e^{\beta \omega'} - 1} 
    \exp[-\im(\omega-\omega')t] + \\%
    &
    \left[\int_{0}^{\infty} \frac{\dd{\omega}}{\pi} J(\omega) \coth{\frac{\beta\omega}{2}}\right]^2.
\end{aligned}
\end{equation}
\end{widetext}
Substitute this result into Eq.~\ref{eqn:FGR-continuous}, we observe time integration $e^{\pm \im \epsilon t}$, along with complex phases in the first two terms, yields delta functions. After frequency integration, this results in a finite rate value. However, the last term in the correlation function contributes an oscillating integral that does not converge to a finite value {(Eq.~\ref{eqn:oscillatory_term} in the main text):}
\begin{equation}\label{eqn:oscillatory_term-app}
    \left[\int_{0}^{\infty} \frac{\dd{\omega}}{\pi} J(\omega) \coth{\frac{\beta\omega}{2}}\right]^2 \int_{-\infty}^{\infty} \dd{t} e^{\pm \im \epsilon t}
\end{equation}
Notably, both $0 \to 1$ and $1 \to 0$ process share the identical oscillating integral with frequency of spin energy difference $\epsilon$. 

Despite Eq.~\ref{eqn:oscillatory_term} yield a oscillating rate, we argue this term can be neglected in the weak coupling limit. In particular, the extended DEOM simulations shows the $\hat{x}_\text{B}^2$-coupling contribution of the rate kernel has free induction decay like feature (FIG.~\ref{fig:kernel-vs-time}). This indicates higher order interactions ignored in FGR relaxation will introduced relaxation to the oscillating rate term. Appendix.~\ref{app:demo_importance_of_decaying} demonstrates that if the relaxation rate $\Gamma \ll \epsilon$, as is the case of weak coupling, Nonetheless, it suffices to use only the first two finite terms representing the FGR rates, given by
\begin{equation}\label{eqn:rate_quad}
    k_{f \gets i}^{(2)} = \frac{4}{\pi} \int_{-\infty}^{\infty} \dd{\omega}
    \frac{J(\omega)}{1 - e^{-\beta \omega}} \frac{J(\omega \pm \epsilon)}{e^{\beta (\omega \pm \epsilon)} - 1}.
\end{equation}
Once again, Eq.~\ref{eqn:rate_quad} suggests the quadratic rates also satisfy the detailed balance. For quadratic system-bath interaction, the general temperature dependency of the $T_1^{-1}$ depends on the specific form of $J(\omega)$. Nevertheless, we can conclude $k_{1 \gets 0}^{(2)} + k_{0 \gets 1}^{(2)} \propto \beta^{-2}$ approximately holds true in the high temperature limit, from $\lim_{\beta \to 0} (1 - e^{-\beta \omega})(e^{\beta (\omega \pm \epsilon)} - 1) \approx \beta^2 \omega (\omega \pm \epsilon)$.

Together, Eqs.~\ref{eqn:rate_linear}~and~\ref{eqn:rate_quad} are the major analytical results of this work. Respectively, these results correspond to the direct (one-phonon) and Raman (two-phonon) relaxation processes commonly referred by the field.

\section{\label{app:demo_importance_of_decaying} Demonstration: The importance of relaxation mechanism to $T_1^{-1}$ rate}
\begin{figure}[htbp]
    \centering
    \includegraphics[scale=0.8]{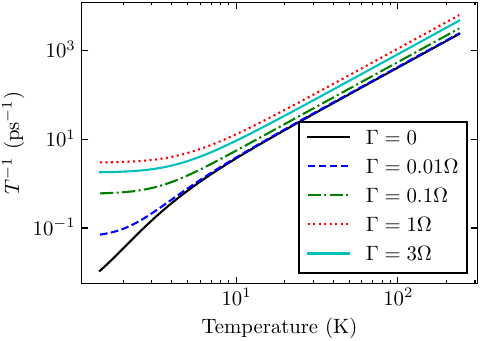}
    \caption{The effect of relaxation parameter $\Gamma$ on the $T^{-1}$ rate. The following parameters are used: $\alpha_0 = 0$, $\Delta = \zeta= \unitval{10/3}{\wavenumber}$, $\omega_\text{B}=\unitval{10}{\wavenumber}$, and temperature $T=\unitval{53.2}{\kelvin}$.}
    \label{fig:effect_of_Gamma}
\end{figure}
The free induction decay like feature of the rate kernel can be modeled by the following equation
\begin{equation}
    f(t) = e^{-\Gamma t} \cos(\Omega t),
\end{equation}
which has analytical integral: $\int_{0}^{\infty} \dd{t} f(t) = \frac{\Gamma}{\Omega^2 + \Gamma^2}$. Thus, if we consider a relaxation rate to Eq.~\ref{eqn:oscillatory_term}, we would have additional contribution to $T_1^{-1}$,
\begin{equation}\label{eqn:}
    \left[\int_{0}^{\infty} \frac{\dd{\omega}}{\pi} J(\omega) \coth{\frac{\beta\omega}{2}}\right]^2 \frac{2\Gamma}{\Gamma^2 + \epsilon^2}.
\end{equation}
Hence, we can neglect the contribution of Eq.~\ref{eqn:oscillatory_term} to the rate if we have no (FGR) or little (weak coupling regime). Nevertheless, we demonstrate in FIG.~\ref{fig:effect_of_Gamma} that this contribution to the rate cannot be neglected if the relaxation rate $\Gamma$ becomes comparable to $\Omega$. To this end, we argue the FGR treatments for $T_1^{-1}$ rate become inadequate when considerable relaxation is introduced by either stronger coupling and/or a mixture of linear and quadratic interaction.


\bibliography{apssamp}

\end{document}